\documentclass[12pt]{iopart}

\usepackage{subfigure} 
\usepackage{graphicx}

\newsavebox{\tempbox}

\usepackage{iopams}  
\begin{document}

\title[Temperature dependent characterization of optical fibres]{Temperature dependent characterization of optical fibres for distributed temperature sensing in hot geothermal wells}
\author{Thomas Reinsch and Jan Henninges}

\address{Helmholtz Centre Potsdam, GFZ - German Research Centre for Geosciences, Telegrafenberg, 14473 Potsdam, Germany}
\ead{Thomas.Reinsch@gfz-potsdam.de}

\begin{abstract}
This study was performed in order to select a proper fibre for the application of a distributed temperature sensing system within a hot geothermal well in Iceland. Commercially available high temperature graded index fibres have been tested under in-situ temperature conditions. Experiments have been performed with four different polyimide coated fibres, a fibre with an aluminum coating and a fibre with a gold coating. To select a fibre, the relationship between attenuation, temperature, and time has been analyzed together with SEM micrographs. On the basis of these experiments, polyimide fibres have been chosen for utilisation. Further tests in ambient and inert atmosphere have been conducted with two polyimide coated fibres to set an operating temperature limit for these fibres. SEM micrographs, together with coating colour changes have been used to characterize the high temperature performance of the fibres.\\
A novel cable design has been developed, a deployment strategy has been worked out and a suitable well for deployment has been selected.
\end{abstract}
\inpress{\MST}
\vspace{2pc}
\maketitle

\section{Introduction}
Temperature is a key parameter to understanding fundamental processes within a wellbore. Transient temperatures help characterize geothermal reservoirs and design a strategy for the use of a geothermal field and a sustainable production of energy.\\
In the past two decades, distributed temperature sensing (DTS) based on Raman Scattering has increasingly been used for wellbore applications \cite{Hurtig1994,Foerster1997,Henninges2005}. Quasi-continuous temperature profiles can be measured in boreholes by using an optical fibre as sensing element. Since no electronics have to be lowered downhole, DTS is especially suited for high temperature applications.\\
Measuring at high temperatures becomes increasingly important with accessing unconventional oil and gas resources or geothermal reservoirs. For the oil and gas industry, maximum well temperatures can be as high as 340$^{\circ}$C \cite{Williams2000}. In conventional hot geothermal wells, e.g. in high-enthalpy areas in Iceland, temperatures up to 350-380$^{\circ}$C can be reached \cite{Arnorsson1995}. In order to increase the energy output from a single geothermal well by an order of magnitude, current research activities aim at producing geothermal fluids with temperatures up to 550-600$^{\circ}$C from depths of about 4-5~km \cite{Albertsson2003,Fridleifsson2005}.\\
In hydrogen rich environments, optical properties of fibres degrade rapidly due to hydrogen ingression and hydroxyl formation \cite{Williams2000,Smithpeter1999,Normann2001}. Hydrogen ingression and subsequent hydroxyl formation causes strong absorption peaks at relevant wavelengths \cite{Stone1982,Humbach1996,Williams2000}. Increasing differential attenuation between the Stokes and Anti-Stokes band skews temperature measurements over time \cite{Normann2001}. Furthermore, the ingression of hydrogen can lead to a total transmission loss for the optical fibre. To reduce the ingression of hydrogen and the subsequent formation of hydroxyl within the interstices of the silica, hermetic carbon coatings have been used as a hydrogen diffusion barrier. Hydrogen diffusion through carbon coatings, however, can be detected at temperatures above 100$^{\circ}$C \cite{Lemaire07}. Recently, pure silica core fibres have been shown to be well suited for application in hydrogen rich environments \cite{Kaura2008}.\\
Using conventional DTS units, the influence of hydrogen on the measured temperatures can be reduced using an experimental set-up where both ends of the fibre can be accessed. Averaging measurements from both ends alleviates the problem of skewed temperature measurements. Another means to mitigate the effect of hydrogen onto measured temperatures is the dual laser principle where measurements from two different incident wavelengths are used \cite{Lee2005,Suh2008}. If only one end of the fibre can be accessed, the dual laser principle can be used to account for wavelength dependent changes along the attenuation profile.\\
For high temperature applications ($<$ 300$^{\circ}$C), polyimide coated fibres are widely used in wellbore applications. Higher temperatures can be measured with metal coated fibres, but it is well known that these fibres have high attenuation values at low temperatures \cite{Bogatyrev2007}. A further drawback of aluminum coated fibres is additional hydroxyl absorption that can be observed at temperatures of 400$^{\circ}$C \cite{Voloshin2009}.\\
This work has been performed within the framework of the HiTI Project (\textbf{Hi}gh \textbf{T}emperature \textbf{I}nstruments for geophysical reservoir characterization and exploitation) in order to test the applicability of fibre optic temperature measurements at temperatures above 300$^{\circ}$C within a hot geothermal well. HiTI aims to provide geophysical and geochemical sensors and methods for the evaluation of deep geothermal wells up to supercritical conditions of water. Up to now, DTS measurements in geothermal wells are documented up to temperatures of 275$^{\circ}$C for a duration of a few hours \cite{Ikeda2003}.\\
In order to select a proper fibre for deployment, different commercially available graded index fibres have been tested at temperatures up to 400$^{\circ}$C within a three zone tube furnace under laboratory conditions.\\ 
Optical attenuation levels have to remain sufficiently low to allow for fibre optic measurements in wellbore applications. Most available DTS systems have an optical budget of $\approx$~20~dB. Different fibre optic measurement techniques, however, have an even smaller budget. With wells being several kilometers deep, attenuation characteristics have to be rather favorable. Four polyimide coated fibres as well as two metal coated fibres have been tested with focus on additional loss levels of the Stokes signal of a 1064~nm laser. Among the polyimide fibres, a fibre with an additional hermetic carbon coating has been tested. Although the ingression of hydrogen, is likely to occur, as the coating material and the ambient atmosphere might act as sources of hydrogen, this study is focused on microbending losses during temperature cycling experiments. The contribution of hydrogen and hydroxyl to the additional loss level, is assumed to be negligible compared to microbending losses due to thermal stress applied to the fibres.\\
After selecting a fibre, further experiments have been conducted to set an operating limit, an important parameter for the selection of a proper well for deployment.\\

\section{Experiments}
Two different sets of experiments have been conducted. First, temperature cycling experiments have been used to select a fibre for deployment. Second, the selected fibres have been heated in ambient air and inert atmosphere at temperatures of 250$^{\circ}$C, 300$^{\circ}$C and 350$^{\circ}$C for a period of approximately 200~h. SEM micrographs have been used to characterize the performance of the fibres at high temperatures. An attempt was made to use colour changes to estimate the state of degradation and therewith additional loss characteristics of the optical fibre.\\
All tests have been performed with 50/125~$\mu$m graded index multimode fibres.

\subsection{Attenuation changes during temperature cycling}\label{sec:exp:select}
\begin{figure}
	\centering
		\subfigure[Attenuation measurements.]{\label{fig:ofen-otdr}\includegraphics[height=2.6cm, angle=0]{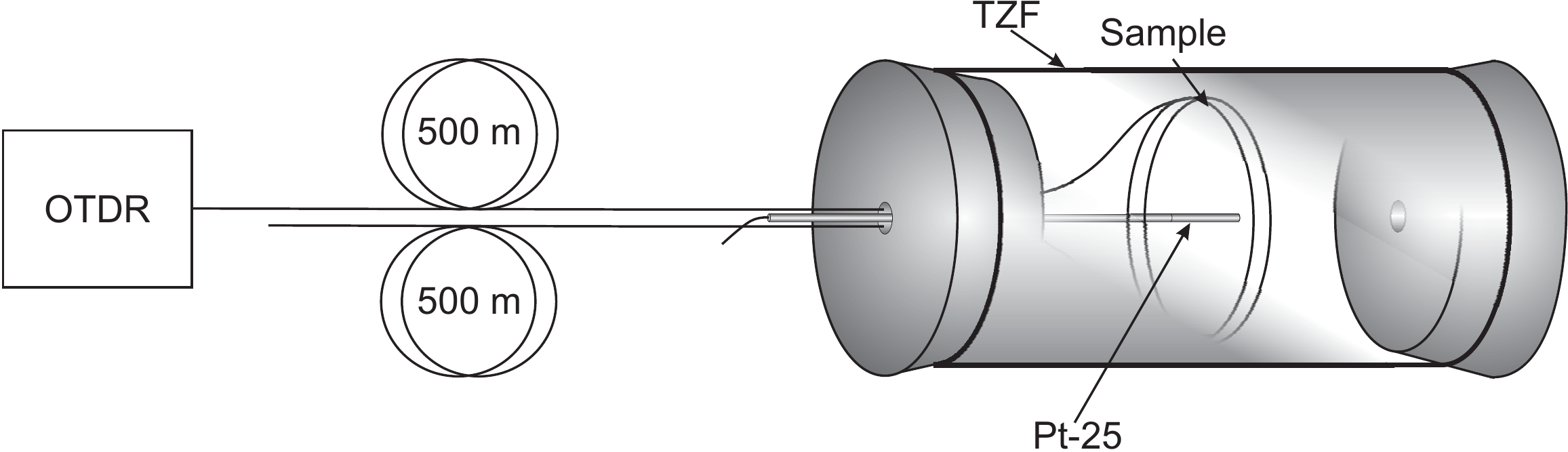}}\hspace{0.1 cm}
		\subfigure[Heating the fibres in ambient air and inert atmosphere.]{\label{fig:ofen_ohneotdr}\includegraphics[height=2.6cm, angle=0]{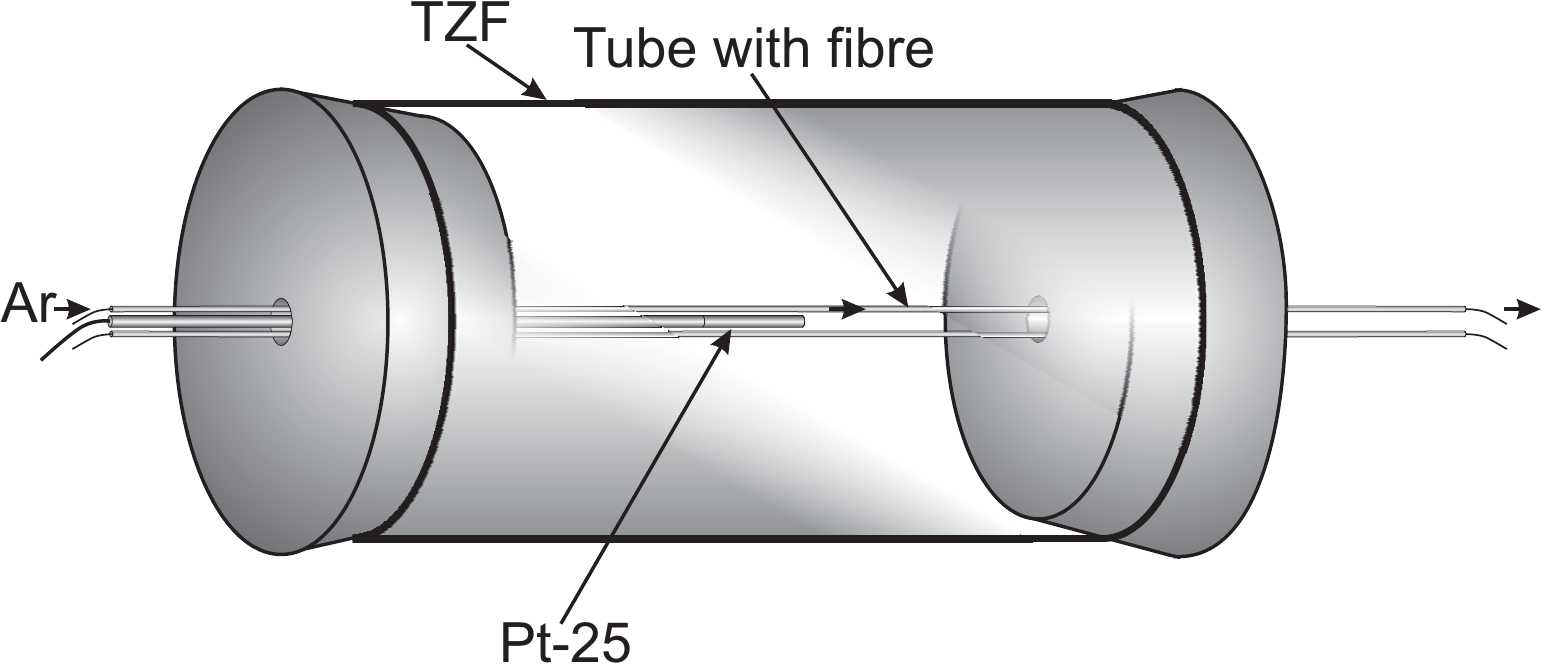}}%
	\caption{(a) Experimental set up for the attenuation measurements. Fibres were spliced to a 500~m launch cable (Section \ref{sec:exp:select}). (b) Experimental set up for heating the fibres in ambient air and inert atmosphere. Small tubes were heated within the TZF and argon was flushed continuously through one of the tubes during heating (Section \ref{subsec:set}). A Pt-25 was used as reference temperature sensor.}
	\label{fig:ofen}
\end{figure}

In order to characterize the different fibres, samples of the fibres  have been heated in ambient air within a Carbolite{\scriptsize$^{\circledR}$\normalsize} three zone tube furnace (TZF) (Figure \ref{fig:ofen-otdr}). 
The samples were rolled and fitted into the tube with a minimum bending diameter of 4.5 cm. A thin sheet of rock wool has been used to separate the fibres from the ceramic wall of the TZF. All samples were spliced to a 500~m launch cable. Attenuation data was calculated from the backscattered Stokes signal of a 1064~nm laser. Data has been acquired using a DTS-800 system from Sensa.\\
Table \ref{tab:list} lists test conditions for different fibres that have been tested within the first set of experiments. Sample lengths have been 20~m for all fibres except the gold fibre (10~m). Temperature cycling has been used to simulate conditions for a wellbore cable that will be subjected to repeated temperature changes, e.g. during logging or well testing operations. To simulate downhole temperature conditions, temperatures were increased for samples 1 to 6 until the maximum testing temperature was reached. Afterwards, the fibres were cooled down to temperatures below 100$^{\circ}$C. Subsequently, fibres 1, 2, 5 and 6 were heated and cooled, again, with a very brief high temperature period for the last cycle. 

\begin{table}

\caption{\label{tab:list}List of tested fibres with respective coating diameters. The maximum operating temperature range (manufacturer specification), the maximum temperature during heating and cooling periods as well as the number of heating and cooling cycles and the total testing time is listed for each fibre. Sample 4 has an additional carbon layer to reduce hydrogen ingression.}

\footnotesize\rm
\centering
\begin{indented}
\item[]\begin{tabular}{cc|cccc}
\br
Sample&Coating&T.-Range&Max.-T.&Cycles&Time\\
&(Diameter~($\mu$m))&($^{\circ}$C)&($^{\circ}$C)&&(h)\\
\mr
1&Polyimide (145)&-90 - 385&385&3&190\\
2&Polyimide (145)&-90 - 385&385&3&190\\
3&Polyimide (155)&-65 - 300&300&1&130\\
4&Polyimide/Carbon (155)&-65 - 300&300&1&130\\
5&Aluminum (155)&-269 - 400&400&4&180\\
6&Gold (155)&-269 - 700&400&4&180\\
\br
\end{tabular}
\end{indented}
\end{table}
\subsection{Heating in inert atmosphere and air}\label{subsec:set}
Polyimide fibres were selected for deployment, on the basis of these experiments. To test the performance of a fibre with an additional carbon layer beneath the coating, and to compare it to a fibre with an ordinary polyimide coating, samples 1 and 4 have been selected for further testing. Samples of 1~m length were heated at constant temperatures of 250$^{\circ}$C for 190~h, 300$^{\circ}$C for 269~h and 350$^{\circ}$C for 216~h, to set an operating limit for these fibres. A sample of each fibre was heated in ambient air and a second one was heated in an argon atmosphere (Figure \ref{fig:ofen_ohneotdr}). Each sample was placed in a stainless steel loose tube with an outer diameter of 2~mm and a wall thickness of 0.2~mm. The tubes were placed in the centre of the TZF and one of the tubes was continuously flushed with Argon 5.0.\\
The effect of argon flow on temperature readings along the fibre is negligible. In order to test this, a 20~m section of a several hundred meter long fibre optic cable was placed in an oven. The temperature difference between inside and outside the oven was 100$^{\circ}$C. Flushing the cable with argon did not change measured DTS temperatures along the fibre.\\
SEM micrographs and coating colour changes have been analysed to estimate the state of degradation. In order to quantify colour changes, high resolution digital photographs have been taken using a Nikon D3 camera. The samples have been illuminated using an Osram Dulux L 18W/12 Lumilux Deluxe Daylight lamp (5400K).

\section{Results}
\subsection{Attenuation changes during temperature cycling}\label{sec:res:sel}
\begin{figure}%
\centering
	\subfigure[Samples 1 and 2.]{\label{fig:PT}\includegraphics[width=0.45\textwidth, angle=0]{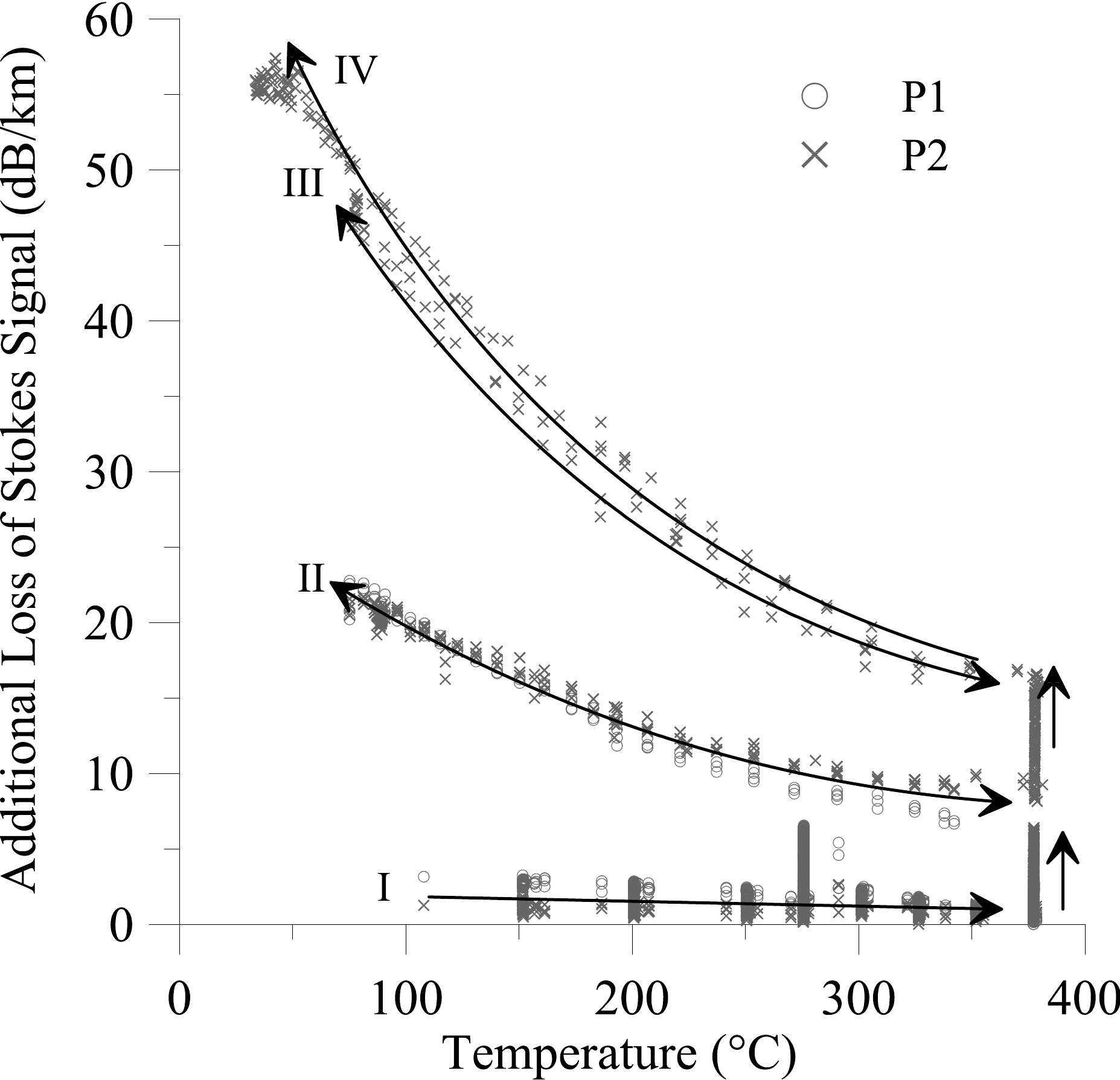}}\quad
	\subfigure[Samples 3 and 4.]{\label{fig:HPPT}\includegraphics[width=0.45\textwidth, angle=0]{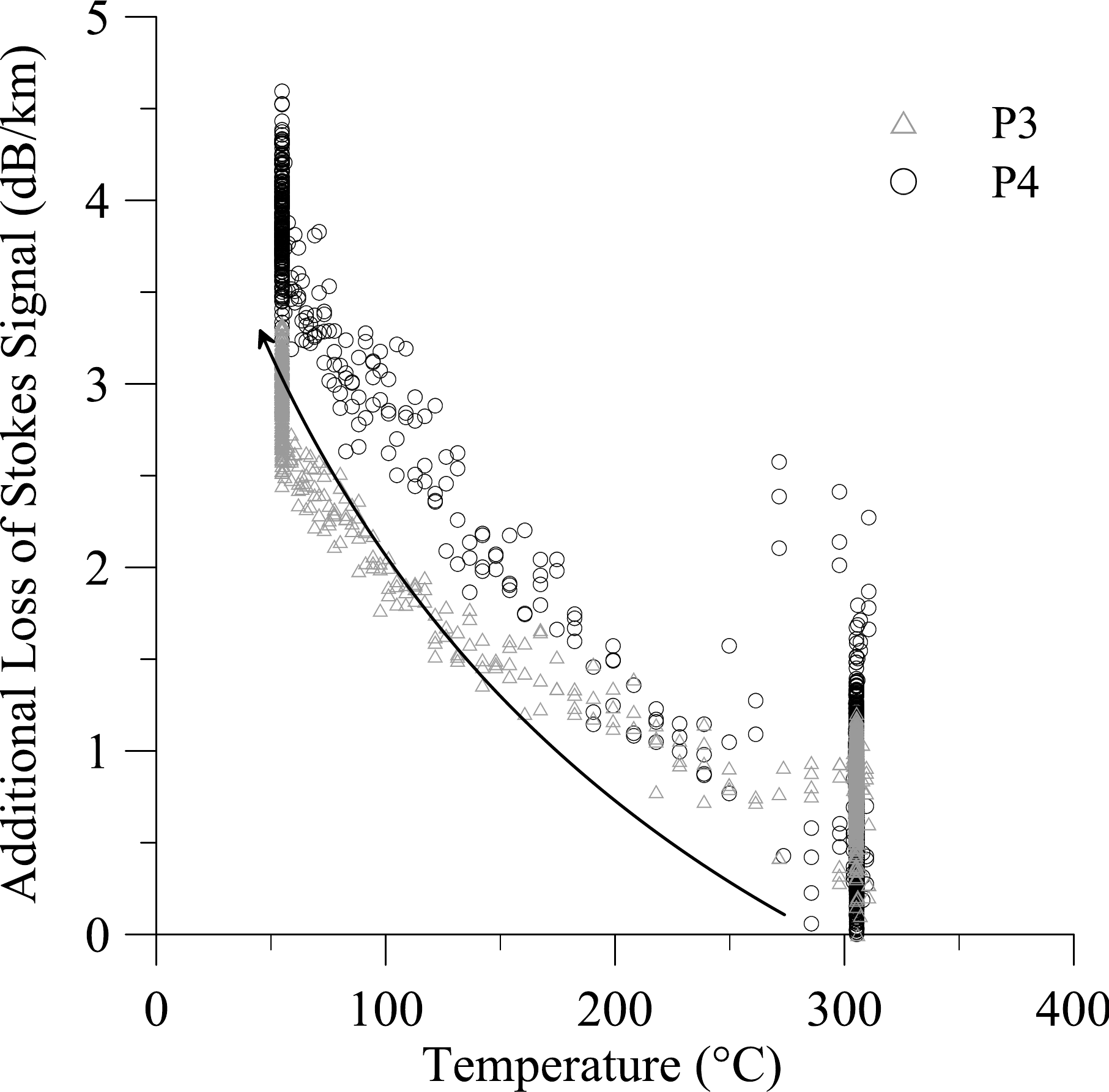}}
\caption{Additional loss vs. temperature for fibre samples 1 - 4. Arrows indicate the evolution in time. Individual heating and cooling cycles are labeled with Roman numbers. Fibre sample 1 broke at the beginning of the second cycle (a). For samples 3 and 4, a single temperatur cycle is displayed (b). Data acquisition started at the end of the heating phase. Note the different scales on the abscissa.}
\end{figure}

\begin{figure}%
\centering
\subfigure[Loss at 300$^{\circ}$C.]{\label{fig:HPP300}\includegraphics[width=0.45\columnwidth]{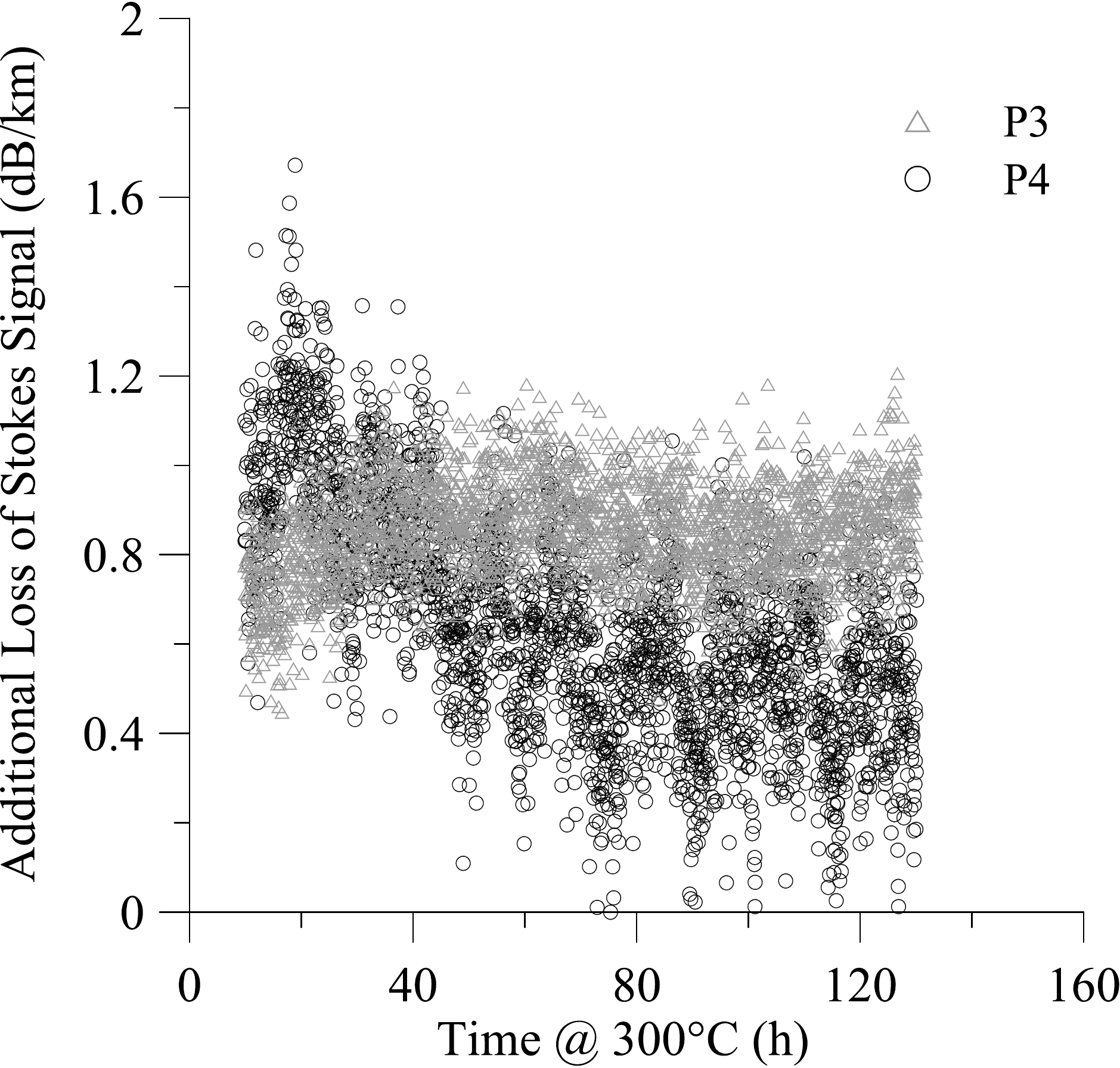}}\quad%
\subfigure[Samples 3 and 4 after heating.]{\label{fig:HPP}\includegraphics[width=0.45\columnwidth]{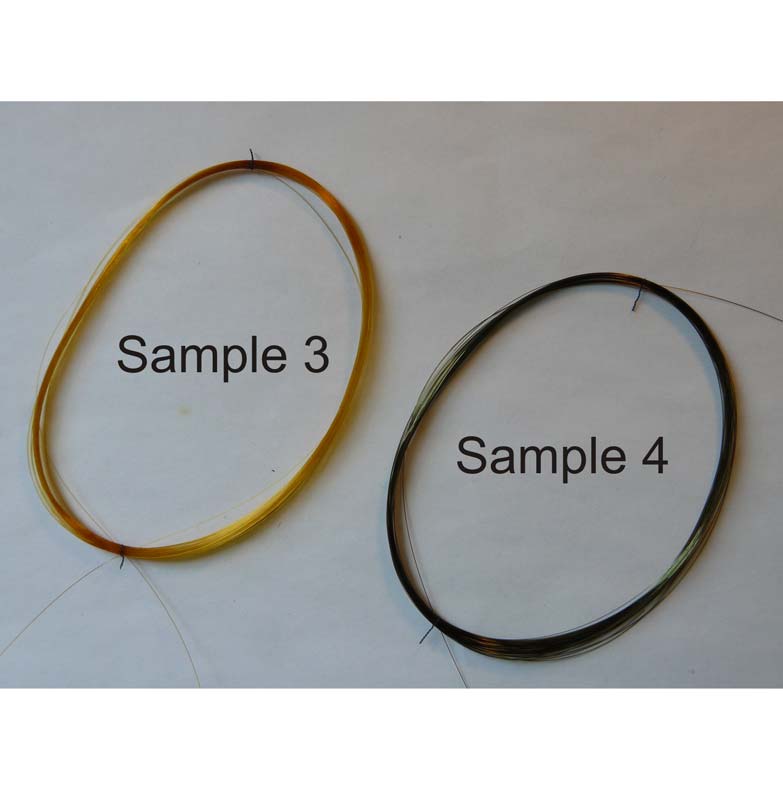}}
\caption{(a) Additional loss vs. time at 300$^{\circ}$C for fibre samples 3 and 4. Attenuation values slowly decrease for sample 4. (b) Picture of fibres after heating.}
\label{fig:HPP-ges}%
\end{figure}

\begin{figure}%
	\centering
		\subfigure[Aluminum.]{\label{fig:AT}\includegraphics[width=0.45\columnwidth]{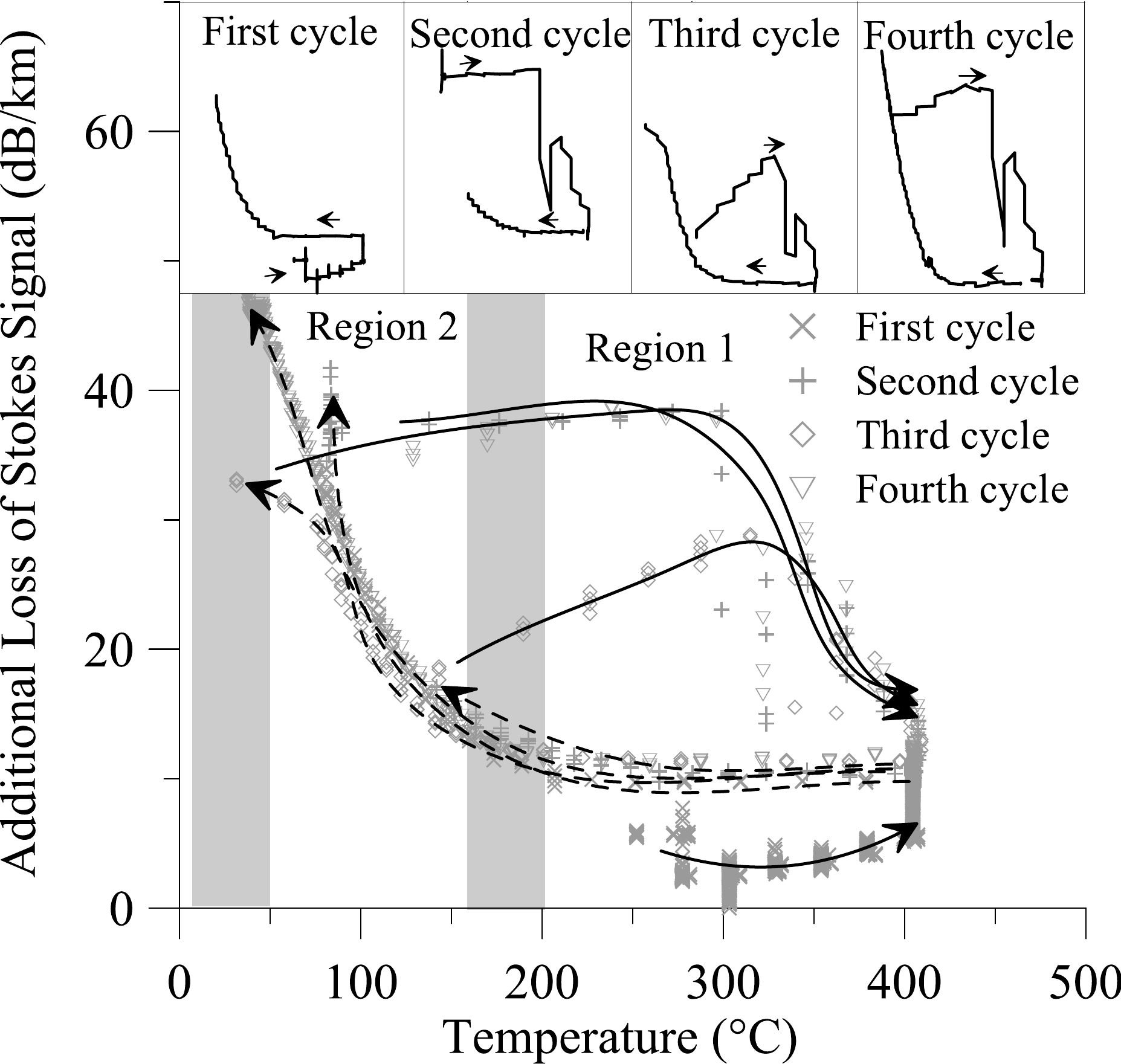}}\quad
		\subfigure[Gold.]{\label{fig:GT}\centering\includegraphics[width=0.45\columnwidth]{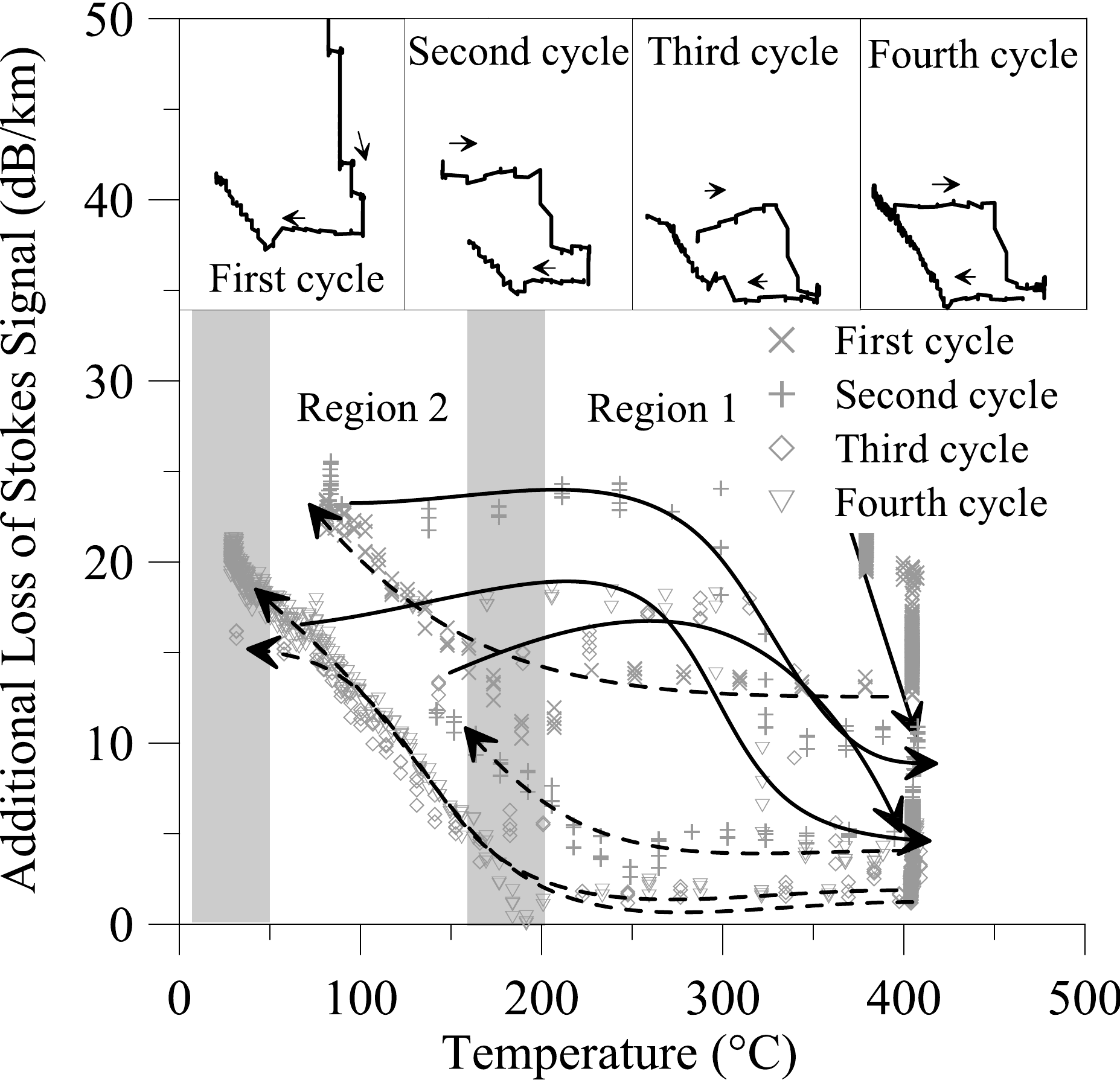}}
	\caption{Additional loss vs. temperature for fibre sample 5 (a) and 6 (b). Arrows indicate the evolution in time. Individual heating and cooling cycles are displayed separately on top.}
	\label{fig:metal}%
\end{figure}

In Figure 2, additional loss versus temperature is shown for the polyimide fibres (samples 1-4). For sample 2, three complete cycles could be measured, whereas only one could be measured for sample 1 due to an early fibre break (Figure \ref{fig:PT}). During the first heating period, attenuation levels stayed rather constant, except for a greater variation around 275$^{\circ}$C. Here, temperature was held constant for a period of 60~h. During the period of constant temperature at 375$^{\circ}$C, attenuation increased irreversibly. During cooling attenuation levels increased reversibly, as attenuation values decreased during subsequent heating. This behavior was observed for all of the three heating and cooling cycles. Reversible attenuation changes occur during heating and cooling; irreversible attenuation changes occur during the 375$^{\circ}$C periods. The slope of the reversible attenuation changes increased in subsequent temperature cycles.\\
Samples 3 and 4 were tested up to 300$^{\circ}$C (Figure \ref{fig:HPPT}). Only one temperature cycle was performed. The fibre samples were held at elevated temperature for more than 120~h. Attenuation values remained constant for sample 3 and decreased at this temperature for sample 4 (Figure \ref{fig:HPP300}). During cooling, attenuation values rose for both fibres (Figure \ref{fig:HPPT}). The slope of this increase was ten times lower than the slope observed for fibres 1 and 2. Taking the fibres out of the TZF, the fibre rolls remained in an oval shape and got stuck to each other (Figure \ref{fig:HPP}).\\
As examples for metal coatings, results for samples 5 and 6 are shown in Figure \ref{fig:metal}. The plots show basically two regions. Low temperature region~2 has high attenuation values, high temperature region~1 has low attenuation values. During temperature cycling, elevated attenuation values decreased at approximately 300$^{\circ}$C during the heating periods. During cooling, attenuation stayed constant until a temperature of approximately 200$^{\circ}$C (Transition Region~1 - Region~2) was reached. Further cooling led to an increase in attenuation values (Region~2). This increase was reversible, and attenuation values decreased during the next heating period, again. Generally, a decrease in attenuation was observed for every heating period. Nevertheless, this decrease was observed at a higher temperature compared to the increase during cooling periods.\\
The overall temperature-attenuation behavior is similar for both fibres. Initial attenuation was low for the aluminum fibre, however, whereas it was high for the gold fibre. Lowest attenuation levels for the gold fibre were measured after the third heating period. Absolute additional loss values are approximately doubled for the aluminum in comparison to the gold coated fibre.

\subsection{Heating in inert atmosphere and air}
\begin{figure}
	\centering

\begin{minipage}{0.25\textwidth}

\centering
Sample 1

\end{minipage}%

\hspace{0.125\textwidth}
\vline
\hspace{0.125\textwidth}

\begin{minipage}{0.25\textwidth}

\centering
Sample 4

\end{minipage}\\%
\begin{minipage}{0.24\textwidth}
	\centering
	Ambient Air
\end{minipage}%
\begin{minipage}{0.25\textwidth}
	\centering
	Argon
\end{minipage}
\begin{minipage}{0.25\textwidth}
	\centering
	Ambient Air
\end{minipage}%
\begin{minipage}{0.24\textwidth}
	\centering
	Argon
\end{minipage}\\
		\subfigure[]{\label{fig:1}\includegraphics[width=0.25\textwidth, angle=0]{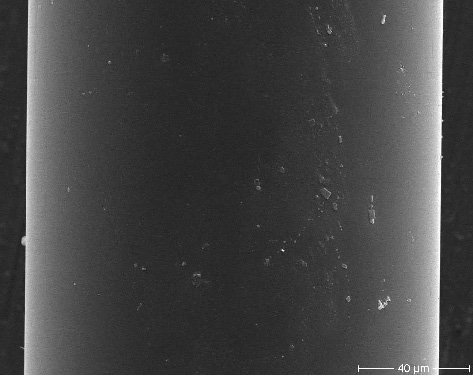}}
		\subfigure[]{\label{fig:2}\includegraphics[width=0.25\textwidth, angle=0]{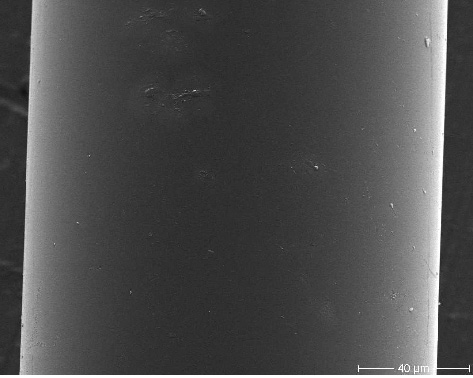}}
		\subfigure[]{\label{fig:3}\includegraphics[width=0.25\textwidth, angle=0]{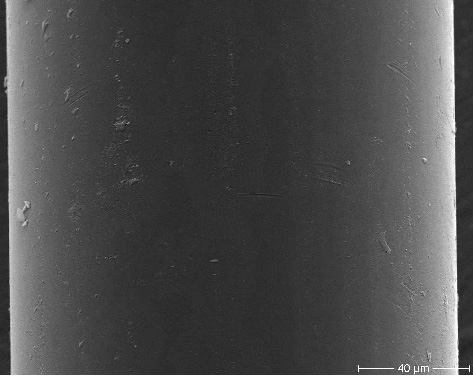}}
		\subfigure[]{\label{fig:4}\includegraphics[width=0.25\textwidth, angle=0]{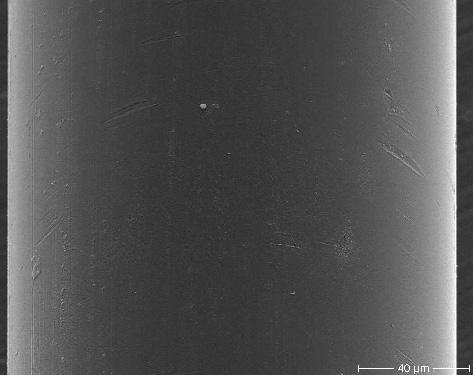}}\\
		\subfigure[]{\label{fig:5}\includegraphics[width=0.25\textwidth, angle=0]{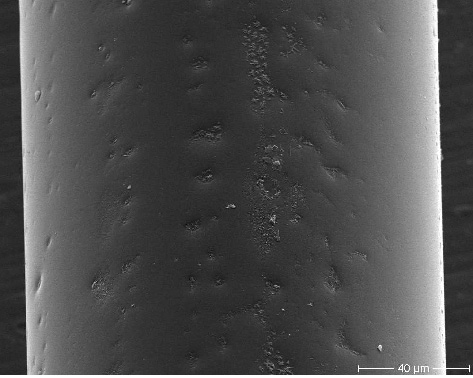}}
		\subfigure[]{\label{fig:6}\includegraphics[width=0.25\textwidth, angle=0]{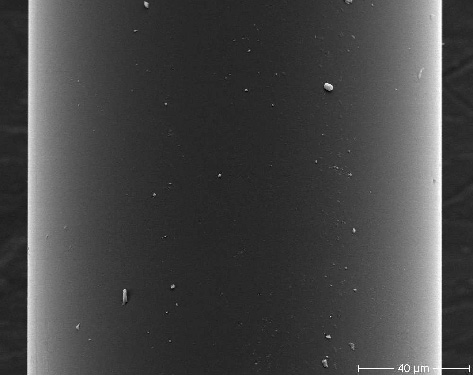}}
		\subfigure[]{\label{fig:7}\includegraphics[width=0.25\textwidth, angle=0]{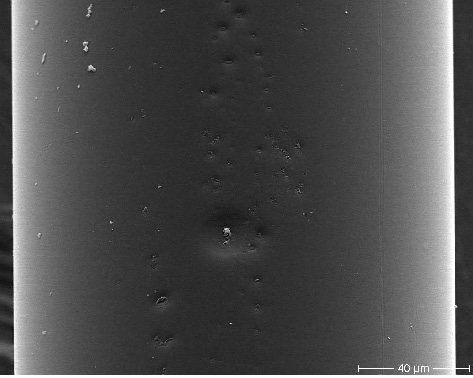}}
		\subfigure[]{\label{fig:8}\includegraphics[width=0.25\textwidth, angle=0]{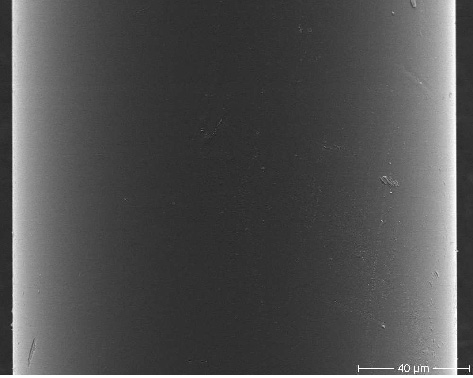}}\\
		\subfigure[]{\label{fig:9}\includegraphics[width=0.25\textwidth, angle=0]{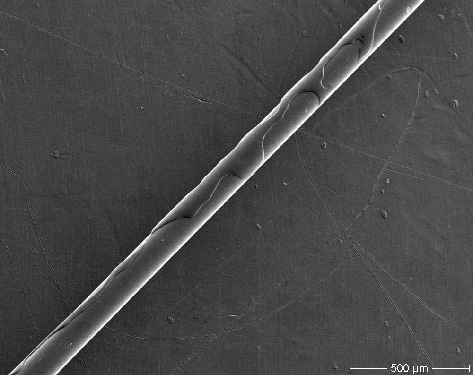}}
		\subfigure[]{\label{fig:10}\includegraphics[width=0.25\textwidth, angle=0]{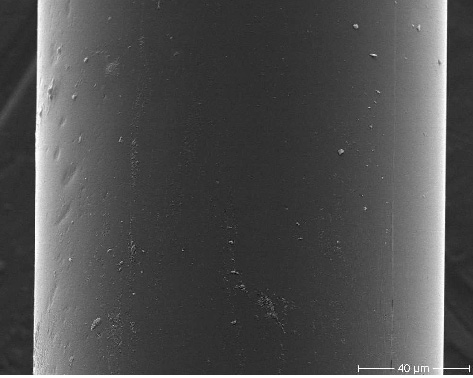}}
		\subfigure[]{\label{fig:11}\includegraphics[width=0.25\textwidth, angle=0]{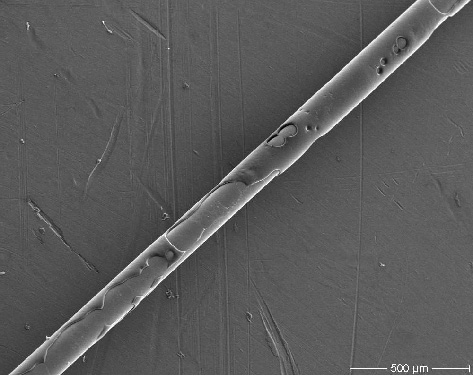}}
		\subfigure[]{\label{fig:12}\includegraphics[width=0.25\textwidth, angle=0]{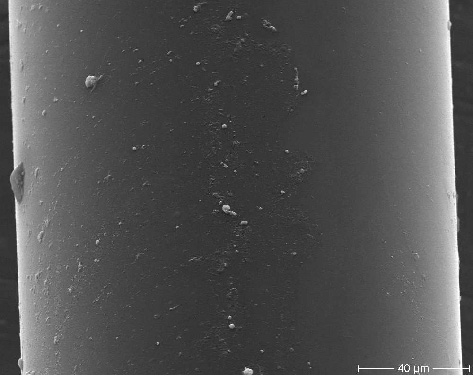}}
	\caption{SEM micrographs from fibres 1 and 4. Samples were heated at constant temperature. The scale bar in the lower right corner indicates 40~$\mu$m for all pictures except (i) and (k) where it is 500~$\mu$m. (a)-(d): 250$^{\circ}$C, 190~h; (e)-(h): 300$^{\circ}$C, 269~h; (i)-(l): 350$^{\circ}$C, 216~h.}
	\label{fig:select}
\end{figure}

Figure \ref{fig:select} shows SEM micrographs of fibre samples 1 and 4 heated at different temperatures in ambient air and argon atmosphere. SEM micrographs of the fibres clearly show an increase in size and frequency of surface depressions with temperature. Furthermore, depressions increase both in size and frequency in ambient air, compared to fibres heated in argon atmosphere. The depressions were not evenly distributed along the fibre. Sample 4 developed fewer and smaller depressions compared to sample 1.

\subsubsection{Colour Changes} 
\begin{figure}%
	\centering
		\subfigure[P1.]{\label{fig:cc1}\includegraphics[width=0.45\columnwidth]{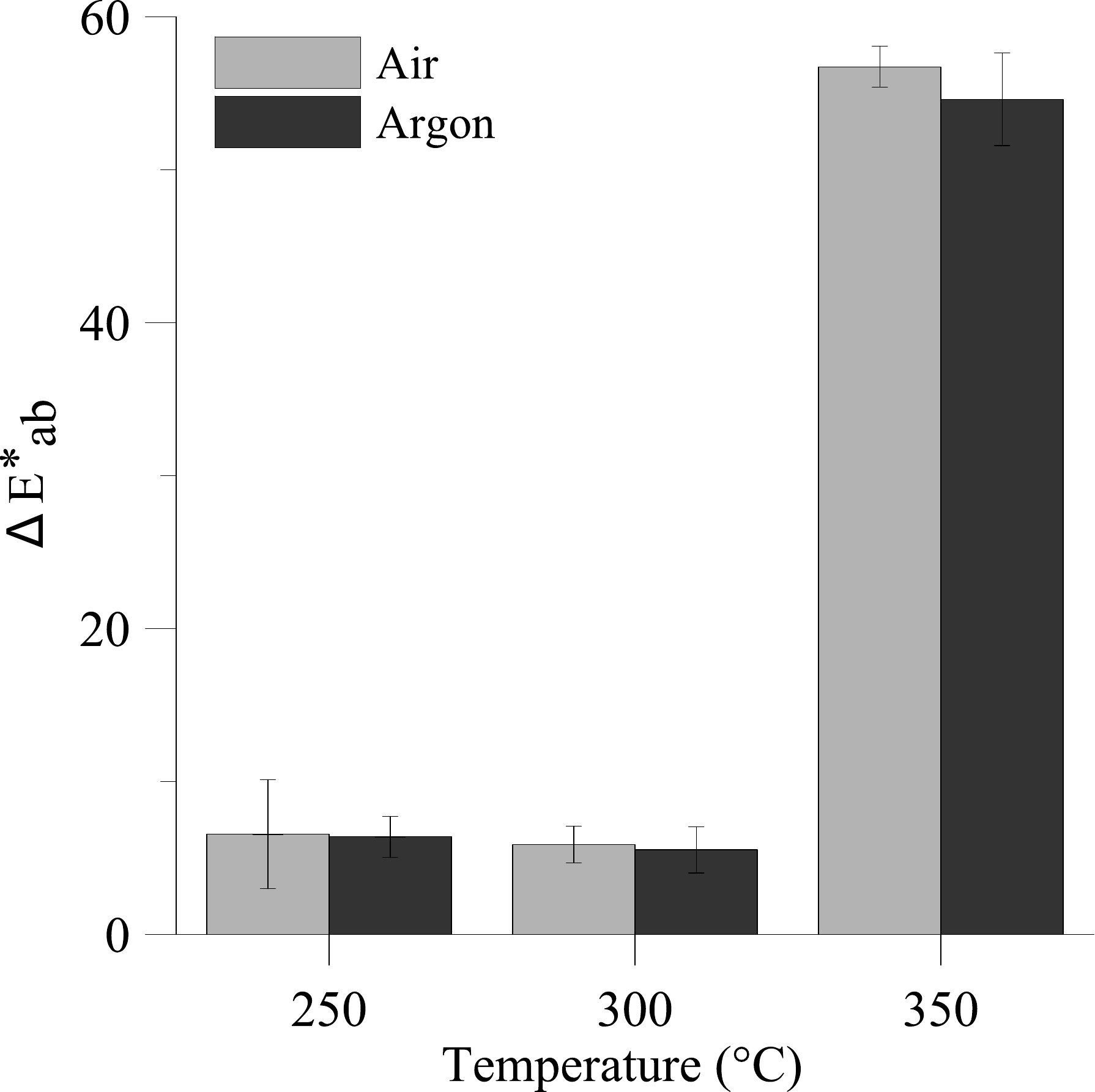}}\quad
		\subfigure[P4.]{\label{fig:cc2}\centering\includegraphics[width=0.45\columnwidth]{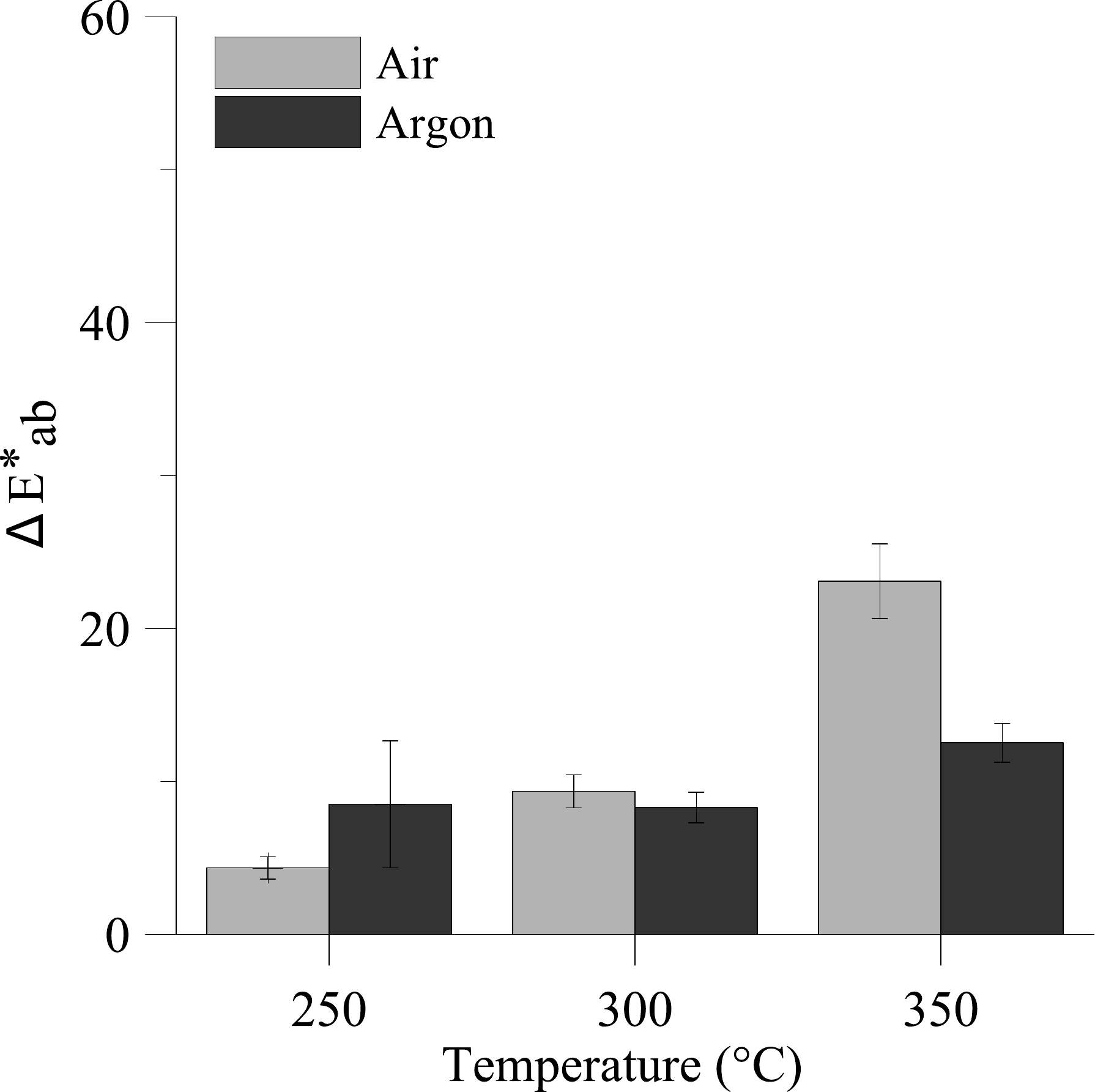}}
	\caption{Colour change $\Delta E^{\ast}_{ab}$ for sample 1 and 4. Non heated fibre samples have been used as reference.}
	\label{fig:cc}%
\end{figure}

Due to degradation, the cladding was partly exposed for samples heated at 350$^{\circ}$C. For each fibre, three representative areas where the coating was still covering the cladding have been used to determine a sRGB value for the coating colour. These values have been converted to the CIE 1976 L$^{\ast}$a$^{\ast}$b$^{\ast}$ colour space \cite{Cie2004} using a CIE 1931 - 2$^{\circ}$ standard colorimetric observer. The L$^{\ast}$ axis represents colours from black to white, a$^{\ast}$ represents green to red and b$^{\ast}$ blue to yellow. Averages and standard deviations of the colours has been calculated for each fibre from the recorded digital photographs. The colour difference along a single axis has been calculated as the norm $\Delta L^{\ast}=\left|L^{\ast}-L^{\ast}_{R}\right|$, $\Delta a^{\ast}=\left|a^{\ast}-a^{\ast}_{R}\right|$ and $\Delta b^{\ast}=\left|b^{\ast}-b^{\ast}_{R}\right|$, respectively. The subscript $R$ denotes the reference sample. Total colour differences $\Delta E^{\ast}_{ab}$ have been calculated as the Euclidean distance within the L$^{\ast}$a$^{\ast}$b$^{\ast}$ space according to equation \ref{eq:e}.
\begin{equation}
\Delta E^{\ast}_{ab}=\sqrt{(\Delta L^{\ast})^{2}+(\Delta a^{\ast})^{2}+(\Delta b^{\ast})^{2}}
\label{eq:e}
\end{equation}
The calculations are based on colorimetry and the results are therefore illuminant-dependent. In order to compare absolute values of colour differences, measurement conditions have to be identical.\\ 
Colour changes were observed for all fibres that have been heated (Figure \ref{fig:cc}). Different fibres show different colour changes with temperature and atmosphere. The colour remains similar, however, for the samples heated at 250$^{\circ}$C and 300$^{\circ}$C. $\Delta E^{\ast}_{ab}$ values stay below 10 compared to the reference fibre. At 350$^{\circ}$C colour changes considerably for all fibres and $\Delta E^{\ast}_{ab}$ increases to more than 10.\\
Initially, polyimide fibres have a light yellow colour. The additional carbon layer beneath the polyimide coating of sample 4 causes a dark, greenish appearance of this fibre. After heating, the colour appearance did not change for sample 4 as much as for the light yellow sample 1 without the carbon layer. The difference at 350$^{\circ}$C  in comparison to the colour changes at 300$^{\circ}$C is large for fibre sample~1, whereas it appears to be small for sample~4. 

\section{Discussion}
\subsection{Attenuation changes during temperature cycling}\label{sec:dis:sel}
\paragraph{Polyimide coated fibres}
At elevated temperatures, additional loss for samples 1 and 2 shows an irreversible increase in attenuation levels (Figure \ref{fig:PT}). This is in good agreement with the fact that polyimide coatings suffer from curing, thermal, thermoxidative and moisture induced degradation at elevated temperatures \cite{Cella1996,Stolov2008}. Degradation rates increase with increasing temperature. The material looses plasticity and stress onto the silica structure of core and cladding increase irreversibly. Hence, microbending loss increases attenuation levels \cite{Williams2000}.\\
During cooling periods, attenuation changes reversibly. After initial heating at elevated temperatures, a more rigid coating and different coefficients of thermal expansion (CTE) cause an increase of stress onto the silica during cooling. Depending on the chemical composition, the CTE for polyimide (e.g. $34*10^{-6}/K$ (30-400$^{\circ}$C) \cite{DuPont2008}) is roughly one to two orders of magnitude larger than the CTE of silica ($0.5*10^{-6}/K$ \cite{Roy1989}). This effect is reversible, and stress is released during subsequent heating cycles. The slope of the reversible attenuation changes during subsequent temperature cycles increases with reduced elasticity of the coating material.\\
Due to considerably increased degradation rates at temperatures above 300$^{\circ}$C, attenuation values increase irreversibly for samples 1 and 2. At constant temperatures around 300$^{\circ}$C, attenuation stays constant for sample 3, and decreases for sample 4 (Figure \ref{fig:HPP300}). Although the oval shape of the fibre roll and the adherence indicate a degradation of the coating material (Figure \ref{fig:HPP}), irreversible attenuation changes have not been observed at this temperature. Attenuation changes during cooling indicate a rigidification of the polyimide coating at 300$^{\circ}$C.\\
\paragraph{Metal-coated fibres}
Metal-coated fibres show high loss values at temperatures below 200$^{\circ}$C (Region~2, Figure \ref{fig:metal}) and low attenuation at temperatures above 200$^{\circ}$C (Region~1). Accelerated plastic deformation of the metal at temperatures above 200$^{\circ}$C and annealing of microbends reduce the attenuation levels in region~1. Plastic deformation is caused by recrystallization processes within the metal and leads to a reduction of residual stresses within the coating. In region~2, slow plastic deformation processes and different coefficients of thermal expansion between silica and metal cause a significant increase in attenuation \cite{Bogatyrev2007}. The transition between Region~1 and 2 was determined to start below 200$^{\circ}$C during cooling. This is in good agreement with results shown in \cite{Bogatyrev2007} for the aluminum coating, whereas it is about 50$^{\circ}$C higher for the gold coating.\\
The reduction of additional loss during heating was observed to be at a higher temperature as the loss increase during cooling. Since it was not possible to actively cool down the TZF, temperature decrease during cooling took much longer than temperature increase during heating. Thus, the decrease in attenuation during heating was delayed and the transition would have most probably occurred at similar temperatures compared to the cooling phase.\\ 
An increase of more than 20~dB/km was observed for the tested wavelength in region~1. Absolute additional loss values are almost doubled for aluminum in comparison to gold coated fibres. This is in good agreement with the CTE difference of the materials (Gold: $14.9*10^{-6}/K$, Aluminum: $26.4*10^{-6}/K$ (0-400$^{\circ}$C) \cite{Nix1941}). As the coating diameter is similar for both fibres, absolute additional loss values are mainly influenced by the CTE. The higher the CTE, the more microbending occurs and attenuation increases. Different initial attenuation values might be explained by a different mechanical and thermal stress history, different drawing techniques and different elastic properties of gold in comparison to aluminum.\\
The small bending diameter of 4.5~cm increased microbending effects and absolute loss values are certainly larger than values acquired for fibres heated with a larger bending diameter. Furthermore, the small length of the samples adds an error to the absolute loss values. The general tendency, however, should be similar.\\

\subsection{Heating in inert atmosphere and air}
The SEM micrographs show large differences for fibres heated in ambient air and argon atmosphere (Figure \ref{fig:select}). Fibres heated in argon atmosphere show less signs of degradation as the fibre samples heated in ambient air. Thermooxidative and moisture induced degradation are unlikely to occur due to the low oxygen and moisture content within the flushed tubes. Thermal degradation and curing reactions remain as the most relevant degradation mechanisms (see Section \ref{sec:dis:sel}).\\
Morphological and colour changes were observed to increase with temperature for all samples. This is in good agreement with increasing degradation rates at elevated temperatures \cite{Cella1996,Stolov2008}.\\
Heated in argon atmosphere, sample 4 shows fewer and smaller depressions on the surface after the 350$^{\circ}$C period. Differences in the chemical composition and thickness of the polyimide material might cause this phenomenon. Depressions, however, were not evenly distributed. Either a larger volume would need to be analyzed or themogravimetric measurements would need to be done in order to quantify differences in the volume change and differences in weight due to heating.

\subsubsection{Colour Changes} Observing colour changes is a quick and inexpensive means to approximate the state of degradation of the polyimide coating material. For fibres heated at temperatures up to 300$^{\circ}$C, an irreversible increase in attenuation levels has not been observed (Section \ref{sec:res:sel}) and colour changes remain below $\Delta E^{\ast}_{ab}\approx 10$ (Figure \ref{fig:cc}). For fibre samples heated at temperatures up to 350$^{\circ}$C, colour changes of $\Delta E^{\ast}_{ab} > 10$ have been observed. Attenuation measurements for samples~1 and 2 heated up to 375$^{\circ}$C for a much shorter period of time, show an irreversible increase in loss characteristics. The increase of $\Delta E^{\ast}_{ab}$ seems to correlate with the transition of reversible to irreversible attenuation changes. A more detailed analysis including simultaneous measurements of additional loss and colour changes would need to be done to define appropriate threshold values for every fibre.\\
The additional carbon layer of sample 4 adds difficulty to this simple approach. For sample 1, the colour of the fibre is only determined by the colour of the polyimide material. For sample 4, the colour appears darker due to the underlying black carbon layer. As stated earlier, the polyimide layer degrades at elevated temperature. Eventually, it exposes the underlying cladding in the case of sample 1 or the underlying carbon coating for sample 4. If it comes into contact with air at elevated temperatures, the carbon coating degrades as well. Loss of the black carbon coating significantly increases calculated colour changes for sample 4. Since only areas of the samples, where the polyimide coating covers the fibre, have been considered, this effect should be negligible. 

\section{Conclusion and Outlook}
For high temperature applications, the selection of a proper fibre is essential for a successful and long-lasting installation of the sensing cable. Therefore, different fibres have been tested in order to select a suitable fibre for the deployment in the harsh environment of a geothermal well.\\
One of the limitations for measuring temperatures over distances of several kilometers is the optical budget of the surface readout unit. For most units available today, this budget is about 20-25~dB. For fibres having high attenuation values, the optical budget limits the deployable length. Attenuation values higher than a few dB/km are not tolerated for wellbore applications with sensor lengths of several kilometers. Due to the high attenuation values at low temperatures, the tested metal coated fibres cannot be used for DTS in wellbore applications. Thus, polyimide fibres have been chosen despite their limited operating temperature range in comparison with metal coated fibres. To test the performance of a carbon coated fibre that reduces the hydrogen ingression within the hostile environment of a geothermal well, tests with sample 4 and reference sample 1 have been conducted. To keep attenuation at moderate levels, degradation of the coating material has to be minimized.\\
Further testing has been done in order to select an operating limit for the selected polyimide fibres. If a polyimide fibre is heated in inert atmosphere, degradation of the coating material, especially thermooxidative and moisture induced degradation, can be greatly reduced. Although 350$^{\circ}$C is close to the glass transition temperature of polyimide, utilization at these temperatures might be possible for an extended period of time. Sample 4, in particular, might be suitable for a deployment at temperatures close to 350$^{\circ}$C.\\
The results for $\Delta E^{\ast}_{ab}$ show that colour changes might be used to indicate irreversible attenuation changes at temperatures above 300$^{\circ}$C. An appropriate threshold value, however, would need to be determined for every fibre.\\
Based on these results, a wellbore cable has been designed, fibre 4 has been incorporated and first measurements under in-situ conditions within a geothermal well in Iceland have been performed. Results of long term monitoring will be analyzed with regard to reversible and irreversible attenuation changes. If the optical properties degrade such that loss characteristic do not allow for DTS measurements but the additional loss is reversible, DTS measurements will be possible as soon as the temperature rises, again. If the loss is irreversible, the fibre is lost for further DTS measurements.

\ack
This work was performed within the framework of the HITI project (http://www.hiti-fp6.eu/) and funded by the European Commission in the 6th Framework Programme, Proposal/Contract no.: 019913. The authors would like to thank J\"org Schr\"otter and Christian Cunow for their support during design and performance of the fibre tests as well as the two anonymous reviewers for editing and reviewing the manuscript.

\section*{References}

\bibliographystyle{unsrt}
\bibliography{ofs}

\clearpage
\section*{Statement of Provenance}
This is an author-created, un-copyedited version of an article accepted for publication in Measurement Science and Technology. IOP Publishing Ltd is not responsible for any errors or omissions in this version of the manuscript or any version derived from it. The definitive publisher-authenticated version is available online at http://dx.doi.org/10.1088/0957-0233/21/9/094022 .

\end{document}